# Voting in the European Union:
# The square root system of Penrose and a critical point

Karol Życzkowski[*]

Institute of Physics, Jagielonian University,
ul. Reymonta 4, 30-059 Kraków, Poland
and
Center for Theoretical Physics, Polish Academy of Sciences
al. Lotników 32/44, 02-668 Warszawa, Poland

Wojciech Słomczyński[♦]

Institute of Mathematics, Jagielonian University,
ul. Reymonta 4, 30-059 Kraków, Poland

Kraków, May 27, 2004

Abstract:
The notion of the voting power is illustrated by examples of the systems of voting in the European Council according to the Treaty of Nice and the more recent proposition of the European Convent. We show that both systems are not representative, in a sense that citizens of different countries have not the same influence for the decision taken by the Council. We present a compromise solution based on the law of Penrose, which states that the weights for each country should be proportional to the *square root* of its population. Analysing the behaviour of the voting power as a function of the quota we discover a critical point, which allows us to propose the value of the quota to be 62%. The system proposed is simple (only one criterion), representative, transparent, effective and objective: it is based on a statistical approach and does not favour nor handicap any European country.

---

[*] karol@tatry.if.uj.edu.pl
[♦] Wojciech.Slomczynski@im.uj.edu.pl

# 1. Introduction

The theory of voting may be considered as a part of the game theory, developed for many decades. In this work we are going to analyse the special case of indirect voting: each citizen of a given country has an influence to elect his representative, who will be taking, on his behalf, the decisions in the body. This particular problem has already been discussed in 1946 by an English psychiatrist and mathematician Lionel S. Penrose[1] (see Pe46 and Pe52) in context of a hypothetical distribution of votes in the UN General Assembly.

The very similar problem has become the subject of a vivid discussion recently, as a part of the debate concerning the enlargement of the European Union and the Draft European Constitution. Majority of experts agree that the voting system established by Treaty of Nice (2001) and the draft Treaty establishing a Constitution for Europe (2003) are not based on solid scientific foundations (L02, BW03a, BW03b, FM03, FPS03, PS03, Pl03, ABF04, BJ04, BW04, FM04b, Pl04). In this work we present a possible compromise solution, based on the square root law of Penrose (Pe46) and several later publications (FM00, Confer4750/00, FM03, Ma03, FM04b, Pl04, PSŻ04, SŻ04a, SŻ04b). Analysing the relation of the voting power of European countries as a function of the quota defining the qualified majority in the ruling body, we observe a critical behaviour. Discovery of such a critical point allows us to propose a complete system of voting in the Council of the European Union in an entirely objective way. It may be derived from an axiom of representativeness, which requires that citizens in any European country should have the same influence for the decision taken in the European Council.

The paper is organised as follows. The first part (Sect. 1.-4.) has an expository nature and describes the problem of voting in the Council of the European Union and of defining the potential voting power of a given country. The systems of qualified majority voting: adopted by the Treaty of Nice and proposed by the European Convention are described and compared in Sect. 5 and 6. In Sect. 7 we analyse in detail the features of the system proposed in the draft of the European Constitution and show that it is not representative. A compromise solution based on the law of Penrose is proposed in Sect. 8, while in Sect. 9 we demonstrate the existence of the critical point which allows us to establish the optimal quota. A statistical analysis of ensembles of fictitious countries with random populations is presented in Sect. 10. The last section of the paper contains concluding remarks.

## 2. European background

*I cannot conceive of the Community
without total parity*

Konrad Adenauer to Jean Monnet[2]

The recent fiasco of the Intergovernmental Conference of the European Union in Brussels has caused many persons in Europe, including known politicians and journalists, to analyse the position to be held by each individual Member State pursuant to the Treaty of Nice versus the draft Treaty establishing a Constitution for Europe, as recently adopted by the European Convention. Unfortunately, most of the opinions voiced have been neither based on sound premises nor built on a good theoretical foundation. Instead, their authors have often been perpetuating myths and telling half-truths. Below we attempt to present a factual description of the situation based on the mathematical research and our own analysis of the issues involved. The Brussels Conference has failed primarily due to disagreements over the distribution of votes in the Council of Ministers. These are likely to form the main obstacle during the current negotiations on the Constitution for Europe being carried out under the Irish Presidency. Therefore, it is even more important for members of both European society and our political elite to have a true picture of the situation which is free from myth.

In general, comments analysing the decision-making procedures within the European Union are based on the following claims:

I. **The influence of a given Member State upon the decisions made by the EU Council is proportional to the number of votes it has in the Council;**

II. **An increase in the number of votes allocated to a given Member State in the Council always causes a growth in its voting power;**

III. **In a system under which the weights assigned to each Member State are proportional to its population, the influence of each citizen of the Union upon the making of decisions is the same.**

IV. **The draft of the European Constitution leads to the system which is democratic and representative.**

V. **It is very difficult to construct an objective voting system which would be simple, representative and efficient.**

In this work, which forms an expanded version of our previously published articles (see PSŻ040 and SŻ04a) we will show that **all these assumptions are false**, and propose an **alternative compromise solution for the distribution of votes in the EU Council**, free of the defects characterising both the Nice and the draft Constitution vote distribution systems.

## 3. Calculation of voting power

Before we compare the decision making methods in the EU Council laid down in the Treaty of Nice (2001) and the draft Treaty establishing a Constitution for Europe (2003), we will first analyse the more general problem of the calculation of the voting power of each participant under qualified majority rules. This type of issue arises in the case of all decision-making bodies where decisions are made according to predetermined rules concerning what constitutes a majority: parliaments, boards of share holders of stock companies, the United Nations Security Council, the Board of Governors of the International Monetary Fund, the Electoral College electing the US Presidents, and finally the EU Council of Ministers. Those participating in such decision-making processes may be natural or legal persons, or groups of persons (e.g. parliamentary caucuses), as well as individual countries.

Let us begin with the simple example of a model parliament in which party **X** has won 55% of the seats and party **Y** the remaining 45% and which makes decisions by simple majority rule (more than 50%). Is the voting power of party **X** bigger by only 10% than that of party **Y**? No, it is not, as with its absolute majority party **X** may vote any bill into law. Therefore, it is obvious that it has full power. Mathematical literature on the subject playfully refers to party **X** as the "dictator", and, by analogy to the bridge player who in a given deal lays down his cards on the table and only passively observes the ensuing game, party **Y** as the "dummy". With this simple example, it is immediately obvious that **Claim I** is false, whereas **Claim II** can be rejected with the use of an only slightly more complex example. Let us assume that the Parliament of a particular country is comprised of four different parties with the proportion of mandates in Tab. 1.

**Tab. 1. Number of Members within a model Parliament**

| Party | At the beginning of the term | After the break-up of party B |
|---|---|---|
| A | 42 | 54 |
| B | 24 | - |
| C | 20 | 26 |
| D | 14 | 20 |

After the elections parties **A** and **C** formed a coalition, while parties **B** and **D** remained in opposition. After some time, half the members of party **B** broke away to join party **A**, and the remaining ones joined, in equal proportions, parties **C** and **D**. Thus, the number of the Members representing party **C** in this Parliament increased from 20 to 26. Did the voting power of this party increase as well? Obviously not, as with more than a 50% majority, party **A** acquired "dictator" status, being able to rule independently, while parties **C** and **D** could only play the role of "dummy", passively watching ensuing events.

Thus, how can one determine the effective voting power held by a given party if such power is not necessarily proportional to the number of votes controlled? The English psychiatrist and mathematician Lionel S. Penrose (see Pe46 and Pe52) had already considered this problem in 1946, although, in fact, he did not analyse the division of power within a Parliament, but a hypothetical distribution of votes in the UN General Assembly. Applying his theory to the example presented above, we can adopt the following line of reasoning: in order to calculate a voting-power index (known as the *Penrose index*), we must calculate the number of ways in which a given party may form a coalition with other parties in order to gain the required parliamentary majority. Such an approach is fully justified as we are not analysing a one-off process of forming a government but instead dealing with a procedure followed in voting on many different bills where a different parliamentary coalition may be formed for each particular vote. It can be shown with precision that the Penrose index is a simple function of a probability that in a hypothetical vote, the vote cast by a particular party may be decisive. Thus, a mathematical model can be built, enabling one to estimate the voting power likely to be held by a given grouping.

The pioneering work done by Penrose in the United Kingdom remained unnoticed and unappreciated until 1965 when the American attorney John F. Banzhaf III independently carried out a similar analysis and published the results in the American periodical: "Rutgers Law Review" (see Ba65). As a result, this method for the calculation of voting power became popular and the duly standardised index proposed by Penrose become known as the *Banzhaf index* β. Obviously, the Banzhaf index for the "dictator" is 100% and for the "dummy" it is equal to 0%. In the case discussed here, initially the Banzhaf index for party **A** is 50% and for each of the parties **B**, **C**, **D**, approximately 16.7%, to change after the break-up of party **B** into 100% for party **A**, which becomes the "dictator", and 0% for parties **C** and **D**, reduced to the role of "dummy".

To describe an algorithm of computing the Banzhaf index let us consider the case of $N$ countries represented in the assembly. Then the number of all possible coalitions equals $2^N$. Assume that $\omega$ is the number of *winning coalitions*, in the sense that they satisfy the qualified majority rule adopted. There exist $2^{N-1}$ different coalitions in which a given country takes part. Let $\omega_x$ denote the number of winning coalitions that include the country $x$. Assuming that all $2^N$ coalitions are equally likely (which follows from the assumptions that the decisions of voters are taken independently and the probability of 'yes' and 'no' decisions are equal), we can compute the probability that $x$ has an influence for the decision taken. This happens, if $x$ is a *critical voter* in a coalition, i.e., the winning coalition (with $x$) ceases to fulfil the majority requirements without $x$. The number of these cases is: $\eta_x = \omega_x - (\omega - \omega_x) = 2\omega_x - \omega$. The *absolute Banzhaf index* is equal to the probability that $x$ is critical: $B_x = \eta_x / 2^{N-1}$. To compare these indices for decision bodies consisting of different number of players, it is convenient to define the *normalised Banzhaf index*: $\beta_x = \eta_x / \Sigma_x \eta_x$. A simple example of such a calculation is provided in Tab. 2.

It is worth to note that the Banzhaf index was already defined and applied by Penrose in 1946. Furthermore, he has shown that the probability $p_x$ that the country $x$ is on the winning side reads:

$$p_x = (\omega_x + [2^{N-1} - (\omega - \omega_x)]) / 2^N = 1/2 + B_x/2 \qquad (1)$$

so it is a function of the absolute Banzhaf index. Observe that even a 'dummy' becomes a member of the winning side with probability one half, but the influence for the decision taken may be measured by the excess of the probability with respect to this threshold. To characterise the *efficiency* of an entire voting system, one defines the *Coleman index* (called also the *power of collectivity to act*) $A$, equal to the probability that a randomly chosen coalition satisfies the qualified majority rule, i.e., $A = \omega/2^N$.

Let us emphasise explicitly that the voting power defined by the Banzhaf index is only intended to serve as a model quantity and exclusively concerns the potential rather than the actual voting power of each player. Although in the mathematical model one assumes that all the coalitions are equally likely, in reality some coalitions are *a priori* more probable than others. Moreover, concerning the voting power in a parliament, the model does not take into consideration the differences in the discipline among the deputies representing different political parties during a particular vote. The model for the calculation of voting power based on the counting up of majority coalitions is applicable to an even greater degree in the case of analysing institutions in which determinations are made by voting with blocks of votes, and in which alliances are not permanent, but change depending upon the nature of the matter under consideration. The Council of Ministers of the European Union constitutes just such a body.

## 4. Voting power in the EU Council

The model presented above may be used for analysing rules governing the taking of decisions in the EU Council. It is important to clearly differentiate here between the *voting weight* of a given country and its potential *voting power*, the latter reflecting the extent to which it may influence decisions taken by the Council when all possible coalitions between different countries are taken into consideration. The voting power depends on the difference between the number of winning coalitions formed with the participation of a particular country and the number of such coalitions formed without it. Expressing these differences in percentages, we obtain indices reflecting the voting power. Depending on the assumptions made, these may be the Banzhaf (see Ba65) or Shapley-Shubik (see SS54)[3] indices. In this article, following the example of most analysts (see BBGW00, FM00, L02, BW03a, BW03b, FM03, Pl03, ABF04, BJ04, BW04, Pl04) we use the Banzhaf index β. It is easy to show that the voting power held by a given country depends not only on its voting weight but also on the distribution of the weights among all the remaining Member States of the European Union. To sum up: **the distinction between the number of votes and their power is of crucial importance for the problem discussed here**.

Often, during discussions on voting power, it is claimed that the Banzhaf index can only reflect the power of a particular country to form a blocking coalition within the EU Council. This claim is also false. In reality, indices reflecting the power to block a decision (i.e. *Coleman preventive power index*) and the power to form a coalition capable of forcing a decision (i.e. *Coleman initiative power index*) are both mutually proportional and depend proportionally on the Banzhaf index. However, the ratios of proportionality depend on the decision rule. Nevertheless, under any decision rule, if one country has more preventive power than another, then it also has more initiative power. So, both phenomena form, simply, the two sides of the same coin.

It has been the case on a number of occasions in the history of the European Communities that the introduced voting system was devised without consultation with experts or an in-depth analysis of the issues involved. The best-known example of this, discussed e.g. by Felsenthal and Machover (see FM00), is provided by the rules governing voting in the first Council of Ministers of the European Economic Community (EEC) which were applicable from 1958 to 1972. During that period, France, Germany and Italy had four votes each on the Council, Belgium and the Netherlands two, and Luxembourg one, while the qualified majority was 12 votes. It might have seemed that Luxembourg was "over-represented" on the Council as, with a population two hundred times smaller than that of Germany, it held as many as 25% of the latter's votes. In reality, Luxembourg played the role of dummy – any potential majority coalition could perfectly well function without its participation and the power of its vote was equal to zero! This result, illustrated in Tab. 2, is evident without any calculations: since all votes of other 5 states are even, the single voice of Luxembourg cannot make the sum of votes equal to 12.

**Tab. 2.** Calculation of the Banzhaf index for the EEC 1958-72: the number of countries: $N = 6$; the total sum of weights allotted = **17**; the quota $q = 12$; the number of possible coalitions $2^6 = 64$; the number of coalitions, in which a given country participates: $= 2^6 / 2 = 32$; number of winning coalitions: $\omega = 14$; the Coleman efficiency index: $A = 14/64 \approx 22\%$.

| Country ($x$) | weights | number of winning coalitions with $x$ | number of winning coalitions without $x$ | the difference | absolute Banzhaf index | normalised Banzhaf index |
|---|---|---|---|---|---|---|
| | | $\omega_x$ | $\omega - \omega_x$ | $\eta_x$ | $\eta_x/2^{N-1}$ | $\beta_x$ |
| Germany | 4 | 12 | 2 | 10 | 5/16 | 5/21 ≈ 0.24 |
| France | 4 | 12 | 2 | 10 | 5/16 | 5/21 ≈ 0.24 |
| Italy | 4 | 12 | 2 | 10 | 5/16 | 5/21 ≈ 0.24 |
| Netherlands | 2 | 10 | 4 | 6 | 3/16 | 3/21 ≈ 0.14 |
| Belgium | 2 | 10 | 4 | 6 | 3/16 | 3/21 ≈ 0.14 |
| Luxembourg | 1 | 7 | 7 | 0 | 0 | 0 |

After the 1972 enlargement of the EEC to include the United Kingdom, Ireland and Denmark, the decision was made to change this state of affairs and Luxembourg was assigned two votes on the Council of Ministers. However, upon the next expansion of the EEC, which brought in Greece, Luxembourg came to match Denmark and Ireland in voting power, although each of the latter two countries had one more vote on the Council.

## 5. Voting rules laid down in the Nice Treaty and the draft Constitution

The EU legislative procedure can be described briefly as follows: legislative proposals are put forward by the European Commission, voted by the European Parliament, and then approved by the Council of Ministers. The forthcoming enlargement of the EU to 25 Member States makes the adoption of new rules governing voting in the enlarged Council necessary. The relevant provisions of the Treaty of Nice and of the draft Treaty establishing a Constitution for Europe differ considerably in this regard.

Pursuant to the Treaty of Nice, votes in the Council are measured by weights which to some degree reflect the population of each individual Member State (see Tab. 2). The Council adopts a piece of legislation if:

(a) The sum of the weights of the individual Member States voting in favour exceeds **232** (with the sum of the weights of 25 Member States being **321**) which is approximately **72%**;

(b) The Member States constituting a qualified majority represent at least **62%** of the total population of the Union;

(c) A majority of Member States (i.e. at least **13** out of **25**) vote in favour.

The outcome of the vote is binding if all three of these conditions are met, but, as the mathematical analysis has shown, condition (a) is the most significant one, as the probability of forming a coalition which would meet only this condition and not meet the other two, is extremely low.

The system for calculating a qualified majority of votes provided for in Article 24(1) of the draft Constitution is based on only two criteria, which must both be met for a piece of legislation to be adopted:

(b) Member States comprising the qualified majority must represent at least **60%** of the total population of the Union;

(c) A majority of Member States (i.e. **13** out of **25**) must vote in favour.

As can be seen, the draft Constitution removes condition (a): adding the weights of the individual votes. To put it differently, weights directly proportional to the population of each individual Member State are applied.

# 6. Which countries gain on the changes proposed by the draft of the European Constitution?

Tab. 3 presents basic data essential for the analysis of both systems for calculating a qualified majority in the Council of Ministers. The table shows the population of each individual Member State[4] (expressed in absolute values and as a percentage of the total population of the European Union of 25), which in accordance with the proposal of the Convention is the only factor determining the weight of their respective vote, as well as the weights of each individual Member State (the number of votes on the Council) as provided for in the Treaty of Nice. The key for comparing both methods is the Banzhaf index β calculated by us[5] for each case.

**Tab. 3. Comparison of voting power (as expressed by the Banzhaf index) of each individual Member State in the EU Council in accordance with the Nice Treaty and the draft Constitution**

| Country | Draft Constitution | | | Nice Treaty | | |
|---|---|---|---|---|---|---|
| | Population (in m) | Population (in %) | Index β (in %) | Votes in the Council | Weight | Index β (in %) |
| Germany | 82,54 | 18,19 | 13,36 | 29 | 9.03 | 8,56 |
| France | 59,63 | 13,14 | 9,49 | 29 | 9.03 | 8,56 |
| United Kingdom | 59,33 | 13,08 | 9,49 | 29 | 9.03 | 8,56 |
| Italy | 57,32 | 12,63 | 9,18 | 29 | 9.03 | 8,56 |
| Spain | 40,68 | 8,97 | 6,96 | 27 | 8.41 | 8,12 |
| Poland | 38,22 | 8,42 | 6,74 | 27 | 8.41 | 8,12 |
| Netherlands | 16,19 | 3,57 | 3,65 | 13 | 4.05 | 4,23 |
| Greece | 11,02 | 2,43 | 2,96 | 12 | 3.74 | 3,91 |
| Portugal | 10,41 | 2,29 | 2,91 | 12 | 3.74 | 3,91 |
| Belgium | 10,36 | 2,28 | 2,91 | 12 | 3.74 | 3,91 |
| Czech Republic | 10,20 | 2,25 | 2,85 | 12 | 3.74 | 3,91 |
| Hungary | 10,14 | 2,23 | 2,85 | 12 | 3.74 | 3,91 |
| Sweden | 8,94 | 1,97 | 2,73 | 10 | 3.12 | 3,27 |
| Austria | 8,07 | 1,78 | 2,62 | 10 | 3.12 | 3,27 |
| Denmark | 5,38 | 1,19 | 2,27 | 7 | 2.18 | 2,31 |
| Slovak Republic | 5,38 | 1,19 | 2,27 | 7 | 2.18 | 2,31 |
| Finland | 5,21 | 1,15 | 2,22 | 7 | 2.18 | 2,31 |
| Ireland | 3,96 | 0,87 | 2,10 | 7 | 2.18 | 2,31 |
| Lithuania | 3,46 | 0,76 | 2,04 | 7 | 2.18 | 2,31 |
| Latvia | 2,33 | 0,51 | 1,87 | 4 | 1.25 | 1,33 |
| Slovenia | 2,00 | 0,44 | 1,81 | 4 | 1.25 | 1,33 |
| Estonia | 1,36 | 0,30 | 1,75 | 4 | 1.25 | 1,33 |
| Cyprus | 0,72 | 0,16 | 1,69 | 4 | 1.25 | 1,33 |
| Luxembourg | 0,45 | 0,10 | 1,64 | 4 | 1.25 | 1,33 |
| Malta | 0,40 | 0,09 | 1,64 | 3 | 0.93 | 1,00 |
| Total | 453,70 | 100.00 | 100.00 | 321 | 100.00 | 100.00 |

A detailed comparison of the changes in the Banzhaf index β resulting from the change in the method of counting votes from the one adopted in Nice to the one worked out by the Convention is presented in Fig. 1.

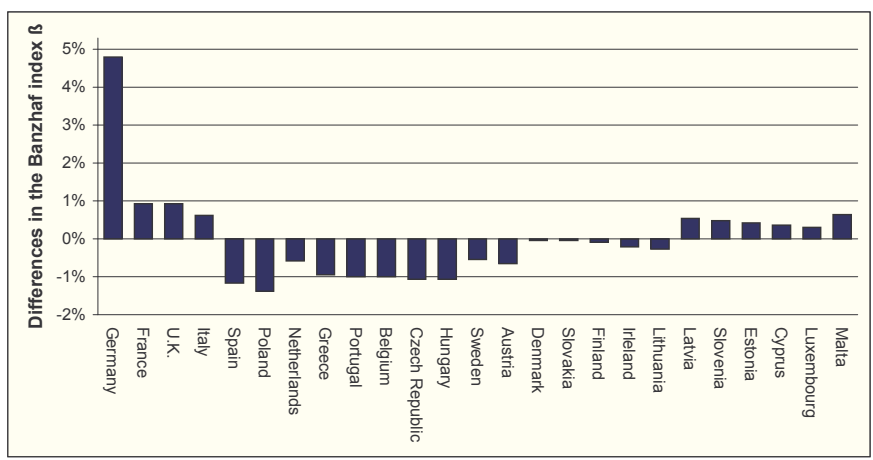

**Fig. 1. Differences in voting power in the EU Council between the proposals put forward by the Convention and the relevant provisions of the Treaty of Nice**

A cursory glance at Fig. 1 provides a quick response to answer the fundamental question: *cui bono*? The most populous countries (Germany, France, United Kingdom, and Italy) gain the most, just due to the demographic factor. The change also benefits the smallest countries (from Latvia to Malta), for which condition (c) is of special relevance. Since the largest and the smallest countries gain voting power, it is easy to see that this occurs at the expense of all the medium-sized countries (from Spain to Lithuania), which loose the voting power. Poland and Spain would lost the most if the rules proposed by the Convention were to be adopted[6].

We decided to examine how the possible conclusion of the Treaty establishing the Constitution for Europe, as drafted by the Convention, would change the influence upon the making of decisions within the Council held by the coalition formed by the EU Member States forming the so-called "inner core", or as some like to put it, the "fast track group": Germany, France, Belgium, and Luxembourg. It is not a simple matter as paradoxically the voting power of coalitions formed by few countries is not equal to the sum of the voting powers of all the members of the coalition. Our calculations indicate that if the rules laid down in Nice were to stand, the potential voting power of this group would be only 14.9%, whereas if the Treaty drafted by the Convention were to be concluded, it would reach 31.2%, increasing by a factor of more than two! No wonder, then, that during the debates within the Convention and the subsequent summit meeting in Brussels these countries fought so strongly for the adoption of rules governing voting in the Council of Ministers which favour them.

Furthermore, it is important to stress that similar results can be achieved by examining how other indices of the potential voting power change (see FPS03).

# 7. Does the draft adopted by the Convention lead to democratic rules of voting?

> […] if two votings were required for every decision, one on a *per capita* basis and the other upon the basis of a single vote for each country, the system would be inaccurate in that it would tend to favour large countries […]
>
> Penrose, 1952[7]

One of the most widely spread myths, cited by various European politicians, says that the voting system in the Council of Ministers of the European Union, as proposed in the draft Treaty establishing a Constitution for Europe, is the simplest and the most democratic possible. Even some of the opponents of the proposals agreed upon by the Convention repeat this claim. Yet it can be proven that the proposed system is not at all the simplest possible and compares unfavourably with the Nice system as regards the realisation of the principle of equality of citizens.

In fact, all voting systems in the decision-making bodies of the European Union, and formerly of the European Communities, have been based on a compromise between two principles: the principle of equality of Member States, and that of equality of citizens. However, even when the emphasis is shifted onto the latter principle, as was the Convention's intent, it is possible to demonstrate with precision that there exists no ideal voting mechanism under which that principle would always be realised.

In indirect elections, the influence of a vote cast by a citizen upon the final-decision-making process depends on the product of the voting power during the direct elections to the government of his country and the voting power of his representative, which in our case is a Member of the Council of Ministers of the European Union. The bigger the population of a particular country, the smaller the voting power of each of its citizens – it is easy to imagine that an average German has smaller influence on the voting of his government than for example a citizen of neighbouring Luxembourg. Assuming for simplicity that the probability of voting 'yes' or 'no' for all citizens in a given country are equal to one half, we need to compute the probability $p$ that half of the constituency of $K$ voters vote 'yes' and half vote 'no'. Making use of the Bernoulli scheme and the Stirling approximation of the binomials one obtains (for even $K$):

$$p = K! / \{2^K [(K/2)!]^2\} \approx [2/\pi K]^{1/2}. \qquad (2)$$

This result, analogous in a sense to the central limit theorem, is valid for a large number of citizens $K$, but this assumption is well satisfied even for the smallest European countries. Hence the voting power of an individual citizen in a direct voting is inversely proportional to the square root of the number of citizens of the country concerned. In order to make up for this interrelation it is necessary, as has already been noted by Penrose, to assign a country participating in the voting a weight roughly proportional to the square root of the number of its citizens – this is the well known *Penrose's Square Root Law*. To put it more precisely, the law proposed by Penrose states that **the influence** **of each citizen of the Union upon the outcome of the voting in the Council will be the same if the voting power of a given Member State in the Council is roughly proportional to the square root of the number of its citizens**, rather than to the number of citizens (*the first square-root rule*, see Pe46, Pe52).

Some people may find this counter-intuitive. To convince them one can use an analogy from physics. In the process of diffusion (or its simplest discrete version: the one-dimensional random walk) the mean square distance of a particle from the origin is proportional to the square root of the time (number of steps). In other words, the diffusion length, i.e., the characteristic length scale for diffusion problems, increases as the square root of the time. From the mathematical point of view this fact follows from the same reasoning as the Penrose Square Root Law. In both cases we use the Stirling formula to estimate, respectively, the probability of a citizen vote being decisive (so votes 'pro' and 'contra' are equally divided) and the probability of the return of a particle to the origin (so 'left' and 'right' steps are equally divided).

There is yet another argument for the application of such weights. Let us examine a simple example of a model "Union" comprised of two Member States: one large State **A** with 36 million citizens and one small State **B** with 4 million citizens. Let us assume that in State **A** 19 million citizens[8] are for the adoption of a particular piece of legislation, and all the citizens of the State **B** are against it. If the weights of the both States are: 36 and 4 respectively, they are proportional to the number of their citizens, and the Ministers vote in accordance with the rules of democracy, i.e. in accordance with the will of the majority in their respective countries, the piece of legislation will be adopted, in spite of the fact that only 19 million of 40 million citizens of the "Union" were in favour of it. What is even more interesting, similar paradoxes can be found for each method for the calculation of qualified majority in indirect voting. In this way it is possible to show that there are no ideal voting arrangements under which the will of the majority of the citizens of the European Union would always prevail. (Under the supposition that all the votes of a Member State in the Council are always cast as a bloc.) On the other hand, one may ask: With what method for the counting of weights to be assigned to each Member State **would the likelihood of a decision being taken against the will of the majority be reduced to a minimum**? It can again be shown in this case that it will be so, **if the voting weight of each Member State in the EU Council will be roughly proportional to the square root of the number of its citizens** (*the second square-root rule*, see FM98, FM99).

These facts are well known to all scientists interested in voting theory (see LW98, BBGW00, L01, W03), but the authors of the draft Treaty establishing a Constitution for Europe ignored these. No wonder than that analysts in many countries are openly sceptical of the arrangements proposed by the Convention (see BW03a, BW03b, FM03, FPS03, PS03, ABF04, BJ04, BW04, FM04b, Pl04, So04), which in addition dramatically shift voting power towards the four most populous Member States. It is most informative in this context to study the book published by Penrose in 1952 (see Pe52). The fragment ending it, quoted at the beginning of this section, may be interpreted as a prophetic and critical opinion on the arrangements laid down in the draft Constitution, formulated 50 years before its adoption by the Convention.

Furthermore, the replacement of the Nice arrangements with the arrangements proposed by the Convention seems to lead to a shake-up of the balance between the three UE bodies: the Commission, the Parliament and the Council. The ease of forming majority coalitions in the UE Council means that its importance decreases in favour of the European Commission, responsible for drafting and putting legislative proposals to the vote, and the European Parliament, responsible for voting on them. Furthermore, it is worthy of note that the weights assigned to individual Member States at Nice are quite close to the Penrose law and therefore **the arrangements adopted at Nice distribute the influence upon the decision-making process in the Council among all the citizens of the Union more evenly than those proposed by the Convention**. The fact that the draft adopted by the Convention makes use of only two criteria, does not remedy another of its basic defects: **an average citizen has no simple way of calculating the potential voting power held by each Member State under this system**; this requires equally complex mathematical calculations as under the Nice system. To sum up arguments presented in this section, we explicitly state that widely accepted **Claims IV** and **V**, presented in the second section of this article, are also false.

Therefore, we have established that the voting arrangements in the Council of Ministers of the European Union laid down in the draft Constitution, are neither advantageous for Poland nor particularly democratic. Seeing this, we may now calmly and somewhat more critically examine the Treaty of Nice. What objective defects characterise the arrangements adopted at Nice? The basic defect lies in the low decision making efficiency. Where in the case of the rules laid down in the draft Constitution, approximately 22.5% of the likely coalitions lead to the adoption of a piece of legislation, in the case of the Nice Treaty this index is only 3.6%, which means that the achievement of qualified majority in the EU Council of Ministers may be difficult under this system$^9$. Another defect of the Nice system lies in the necessity to apply three criteria simultaneously in the calculation of the qualified majority: the weight assigned to each Member State, the number of its population, and the number of the Member States.

The studies conducted by Felsenthal and Machover (see FM01) have shown however that with a slight correction of the weights assigned to each country one can achieve a system exactly equivalent to the Nice one, but which is much simpler, as it is based on the weight criterion alone. Therefore, the rules underlying the Nice system may be cut ruthlessly with Ockham's razor, as the mechanism based on the three criteria is needlessly complex and difficult to comprehend by the average citizen of the Union. On the other hand, the works of Baldwin and Widgrén (see BBGW00, BW03b, BW04) show that with no fundamental change in the voting power of each particular Member State, the Nice system may be modified so that its effectiveness increases many times. One must therefore honestly admit that both defects in the Nice system turn out to be insignificant, while the advantages of the system provided for in the draft Constitution prove to be deceptive.

# 8. Proposed Compromise Solution: P-62 (Jagiellonian Compromise)

It is clear from the preceding analysis that the optimal compromise solution for the problem of voting on the EU Council should satisfy the following conditions:

- The voting system should be as **simple** as possible (optimally, based on a single weighting criterion with a defined threshold), **objective** (that is, founded on scientific principles and not an outcome of political haggling) and easily **modifiable** (so that it can be applied to future enlargement of the EU);

- That system should be moderately **conservative**, that is, it cannot lead to a dramatic transfer of voting power relative to the existing arrangements;
- Weights should be chosen in a way which assures that the system is **representative** (equal influence of each EU citizen on decisions taken by the EU Council);

- The threshold should be set in such a manner that the system be, on the one hand, **effective** (moderately high probability of reaching decisions), and, on the other, **transparent** (potential voting power of a country as measured by the Banzhaf index β needs to be equal to the weight assigned thereto).

Taking advantage of the literature on the subject, we have sought to work out a **compromise solution** that does not duplicate the defects of either the Treaty of Nice or the draft Constitution for Europe, while satisfying **all the above conditions**.

Tab. 4 shows the data calculated for our proposal, which is grounded in Penrose's Square-Root Rule and inspired by the works of Felsenthal and Machover (see Rule B Project, FM00, FM01, FM03, FM04b), as well as referring to a Swedish proposal of 2000 (see CONFER 4750/00, CONFER 4769/00) and similar concepts formulated recently by a French mathematician, Mabille (see Ma03)[10]. The proposed solution (see also SŻ04a and PSŻ04a) is based on two rules, one for the calculation of weights to be assigned to each particular Member State's vote and the other for setting the decision making threshold or quota (the EU Council reaches a decision when the sum of the weights of the Member States voting in favour exceeds that threshold):

1. **In accordance with Penrose's rule, the voting weight of each Member State is allocated proportionally to the square root of its population;**

2. **The decision-taking threshold is set at 62% of the sum of weights of all the EU Member States.**

The above rules, which explain the name[11] given to our proposed solution (**P-62**), require some commentary:

Re. 1: Determining the number of citizens of each Member State based on statistical data[12], requires that some further technical details be worked out. For example, the accuracy of data provided by individual Member States and their sources ought to be assessed, and it is necessary to solve the problem of double citizenship or population migration. Once population data for a given Member State is available, the calculation of the voting weight to be assigned thereto will not be difficult, as a simple calculator with the square root function would suffice for that purpose. The weights set should be adjusted at specified intervals or before each subsequent enlargement of the EU.

Re. 2: The calculation of the value of an appropriate decision-taking threshold is a complex operation (LM03), and its setting materially affects the effectiveness of the system. We have succeeded in identifying a possible objective criterion for choosing such a threshold. As discussed in Sect. 9, its value is calculated in a manner which ensures that the **voting power** of each individual Member State**, as measured by the Banzhaf index** (which also depends on the choice of the threshold), constitutes the best approximation of **the weight proportional to the square root of the population**[13]. This seems to have enormous practical significance: when the threshold is set, the system becomes maximally transparent, for the calculated voting powers correspond almost exactly to the declared weights. Detailed calculations show, moreover, that the threshold set with the use of this method ($R_{opt}$ = 62%) guarantees a reasonable effectiveness of the system.

Furthermore, the **P-62** system (Penrose's rule plus the threshold set at 62%), which represents one of the possible forms of a compromise solution, offers the following advantages:

- It is simple (only a single criterion is required!);

- In contrast to some compromise solutions proposed recently, it has not been produced *ad hoc*, but is based on strict theoretical foundations;

- It may be used in the event of subsequent EU enlargements to include additional countries, such as Bulgaria or Romania, and in the future also Turkey or former Yugoslav countries[14];

- On the one hand, it takes into account the aspirations of Germany, and, on the other hand, while admittedly significantly reducing the voting power of Spain and Poland, it preserves the position of these countries among the "big countries" of the European Union; in contrast to the proposal put forward by the Convention, it does not disturb the existing balance between the Commission, the Council, and the Parliament[15];

- It is representative, as it reflects the principle of equality among all the EU citizens to a significantly greater degree than the proposal put forward by the Convention;

- The decision-making effectiveness index is **16.6%**, which is a much higher value than in the case of the arrangements laid down in the Treaty of Nice;

- The voting power of each Member State, as measured by the Banzhaf index (see Tab. 4, column 6), is almost identical to the weight of the vote cast by its representative on the Council.

**Tab. 4. Comparison of voting power of each particular Member State on the EU Council of Ministers as measured by the Banzhaf index β, in accordance with the P-62 proposal worked out by the authors.**

| Member State | Population (in m) | Population square root | Weight (in %) | Index β (in %) |
|---|---|---|---|---|
| Germany | 82,54 | 9,09 | 10,37 | 10,36 |
| France | 59,63 | 7,72 | 8,81 | 8,82 |
| U.K. | 59,33 | 7,70 | 8,78 | 8,79 |
| Italy | 57,32 | 7,57 | 8,64 | 8,65 |
| Spain | 40,68 | 6,38 | 7,28 | 7,29 |
| Poland | 38,22 | 6,18 | 7,05 | 7,06 |
| Netherlands | 16,19 | 4,02 | 4,59 | 4,59 |
| Greece | 11,02 | 3,32 | 3,79 | 3,79 |
| Portugal | 10,41 | 3,23 | 3,68 | 3,68 |
| Belgium | 10,36 | 3,22 | 3,67 | 3,67 |
| Czech Rep. | 10,20 | 3,19 | 3,64 | 3,64 |
| Hungary | 10,14 | 3,18 | 3,63 | 3,63 |
| Sweden | 8,94 | 2,99 | 3,41 | 3,41 |
| Austria | 8,07 | 2,84 | 3,24 | 3,24 |
| Denmark | 5,38 | 2,32 | 2,65 | 2,65 |
| Slovakia | 5,38 | 2,32 | 2,65 | 2,65 |
| Finland | 5,21 | 2,28 | 2,60 | 2,59 |
| Ireland | 3,96 | 1,99 | 2,27 | 2,27 |
| Lithuania | 3,46 | 1,86 | 2,12 | 2,12 |
| Latvia | 2,33 | 1,53 | 1,75 | 1,75 |
| Slovenia | 2,00 | 1,41 | 1,61 | 1,61 |
| Estonia | 1,36 | 1,17 | 1,33 | 1,33 |
| Cyprus | 0,72 | 0,85 | 0,97 | 0,97 |
| Luxembourg | 0,45 | 0,67 | 0,76 | 0,76 |
| Malta | 0,40 | 0,63 | 0,72 | 0,72 |

Comparing the above data with the Banzhaf indices computed for the two other systems we see that the system proposed has the features of a compromise solution. On one hand the voting power of the states handicapped by the system of Nice (e.g. Germany, Latvia) is larger according to the P-62 system. On the other hand, Spain, Poland and other middle size countries win according to P-62, in respect to the system proposed in the draft of the Constitution. To demonstrate this property we calculate the voting power of an average citizen of a Member State with population $K_i$. It is proportional to the product of the voting power $1/(K_i)^{1/2}$ of a citizen in the local elections, times the voting power of the Minister in the Council (the Banzhaf index $\beta_i$). For simplicity we normalise such an effective voting power with respect to the voting power of a citizen of Germany – the largest country in EU-25. More precisely, we compute the relative quantities

$$\mu_i = \beta_i (K_D)^{1/2} / [\beta_D (K_i)^{1/2}], \qquad (3)$$

where $K_D$ stands for the population of Germany, and $\beta_D$ denotes its Banzhaf index.

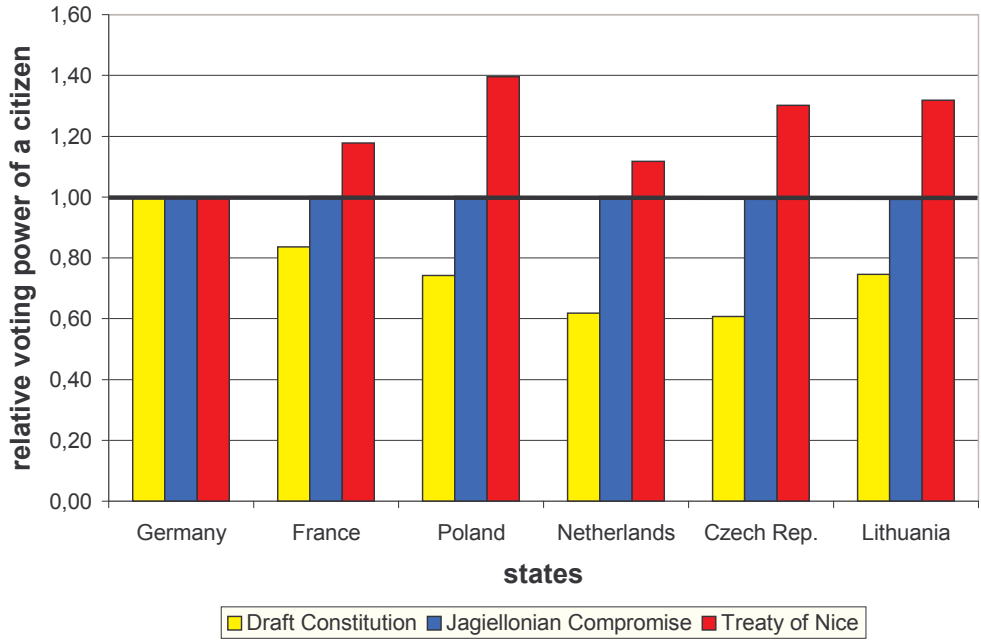

**Fig. 2. Voting power $\mu_i$ of a citizen of France, Poland, Netherlands, Czech Republic and Lithuania relative to the voting power of a citizen of Germany, computed for the Draft of Constitution, the compromise solution P-62, and the Treaty of Nice.**

## 9. A critical point and selection of the optimal decision threshold

The choice of an appropriate decision-taking threshold affects both the distribution of voting power on the Council (and thus also the **representativeness** of the system) and the voting system's **effectiveness** and **transparency**. In working out our compromise solution we have strived to set a threshold that would produce the maximally transparent system, that is, a system under which the voting power of each Member State, as measured by the Banzhaf index, would be equal to their voting weight. Since the weights were chosen to be proportional to the square root of the population (in accordance with the Penrose rule), the resultant threshold simultaneously assures complete representativeness of the system.

Fig. 3 illustrates the dependence of the square root of the sum of square residuals ($\sigma$) between Banzhaf index values and voting weights (for the entire EU) on the value of the threshold ($R$). Since the evident minimum value for this curve is $R_{opt} = 62\%$, we are able to work out the optimal value for the threshold, for which both the voting power and weights coincide.

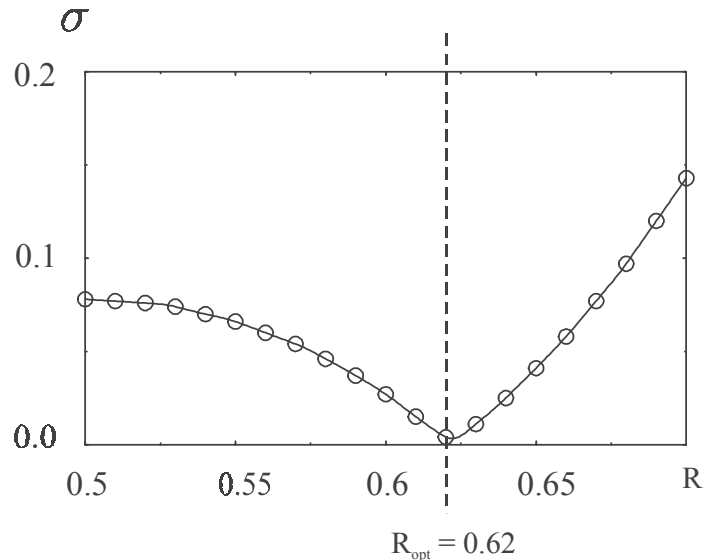

**Fig. 3. Quantity $\sigma$ illustrates the cumulative residual between the voting weight and power for all 25 EU countries as dependent on the value of the threshold $R$**

An alternative method of calculating this value is shown in Fig. 4, which represents the ratio of voting power, as expressed by the Banzhaf index $\beta$, to the voting weight, relative to the threshold value ($R$) for five selected Member States of the European Union. All these curves intersect in the vicinity of the *critical point,* $R_{opt} = 62\%$, for which Banzhaf index values are equal to the voting weights. It is worth noting that, as the threshold is raised, the voting power of the big countries declines while that of the small ones rises. Seen in this perspective, Poland belongs to the group of "middle-sized" countries since its voting power (with the given distribution of weights) appears with good approximation to be independent of the threshold set.

Another argument in favour of the adoption of the threshold $R_{opt} = 62\%$ is provided by an analysis of the effectiveness index. Fig. 5 depicts the dependence of that index value on the threshold variable ($R$) selected. For comparison, it also shows the value of this indicator under the Nice system and the draft Constitution adopted by the Convention. As we can see, when the threshold $R_{opt} = 62\%$ is set, the effectiveness remains at a reasonable level of 16.6%, which exceeds significantly the effectiveness level of 7.8% for the current EU of 15 (see BW03b).

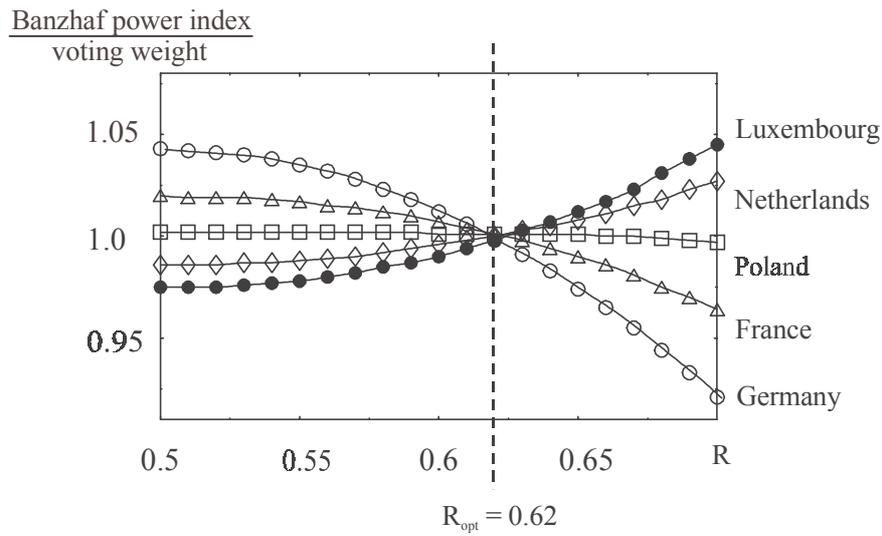

**Fig. 4. Ratio of voting power to vote weight, relative to the threshold set, for five selected Member States (Luxembourg, the Netherlands, Poland, France, and Germany) – all functions cross at the critical point $R_{opt} = 62\%$.**

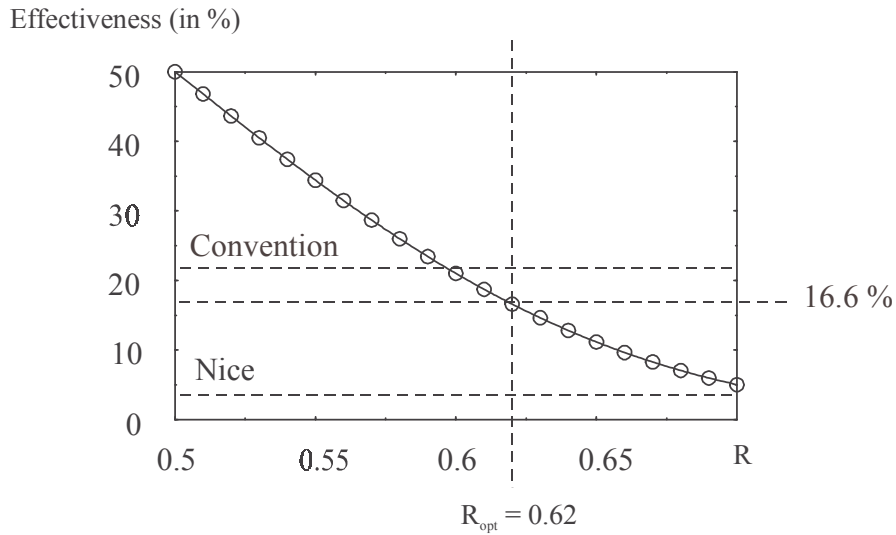

**Fig. 5. Dependence of effectiveness index (Coleman collectivity index) on the threshold value**

## 10. Statistical analysis

From a technical point of view it is worth investigating, whether the critical point observed at the quota $R_{opt} = 62\%$ is characteristic only for a given set of $N = 25$ countries, which form the European Union, or if such a behaviour is generic. To analyse this issue we have repeated the computations described above for a sample of random 'Europe's' consisting of a fixed number of $N$ countries. The total population of the 'Union' has been kept fixed, and the distribution of the population between countries has been drown randomly using two different measures on the probability simplex: a) the *uniform measure* and b) the *statistical* (*Fisher*) *measure*. Both measures belong to the class of the Dirichlet measures defined by the joint probability distribution:

$$p(Z_1, Z_2, ..., Z_N) \sim \delta(1-Z_1-Z_2+...-Z_N) (Z_1 Z_2 ... Z_N)^{s-1} \qquad (4)$$

and correspond to the cases $s = 1$ and $s = 1/2$, respectively. The non-negative numbers $Z_i$ ($1,2,...,N$) denote the fraction of population of the entire 'Union' which inhabits the $i$-th state. In the case $N = 3$ both measures have a nice geometric interpretation: the measure $s = 0$ represents a uniform distribution on the equilateral triangle of probabilities, belonging to the plane $Z_1+Z_2+Z_3 = 1$, while case $s = 1/2$ represents the uniform distribution on the octant of the unit sphere $S^2$.

We have generated randomly ten random fictitious 'European Unions' for each number of states, with respect to both measures, assuming that the number ranges from *12* to *26*. For each realisation we have computed the dependence of the Banzhaf indices on the quota and looked for the critical point. In all cases we have found a deep minimum of the dependence $\sigma(R)$, analogous to this shown for the existing Europe in Fig. 3. The position $R_*$ of the minimum of the mean deviation $<\sigma>$ as a function of $R$ depends only on the measure used and on the number $N$ of the states in the Union. Tab. 5 presents numerical results for the critical threshold $R_*$, under which the deviation of the voting power form the Penrose square root low, are minimal.

**Tab. 5. The optimal threshold $R_*$ obtained from the averaging over ensembles of $N$ states, drown randomly with respect to the Fisher statistical measure (second row), and the uniform measure (third row).**

| N | 10 | 12 | 14 | 16 | 18 | 20 | 22 | 24 | 26 |
|---|---|---|---|---|---|---|---|---|---|
| $R_*(F)$ | 66.0% | 65.8% | 64.6% | 64.4% | 63.4% | 63.1% | 62.6% | 62.0% | 61.4% |
| $R_*(U)$ | 65.5% | 65.2% | 63.6% | 63.2% | 62.9% | 62.2% | 61.7% | 61.2% | 60.6% |

The values of the optimal threshold $R_*$ do not depend strongly on the measure used. Therefore, its choice is not so important for the interpretation of the results obtained. They seem to have a simple practical meaning: Choosing for a given number of players $N$ the quota $R_{opt}$ in vicinity of $R_*$ we assure that the system is almost optimally representative, since according to the law of Penrose, the voting power is proportional to the square root of the population in a given country.

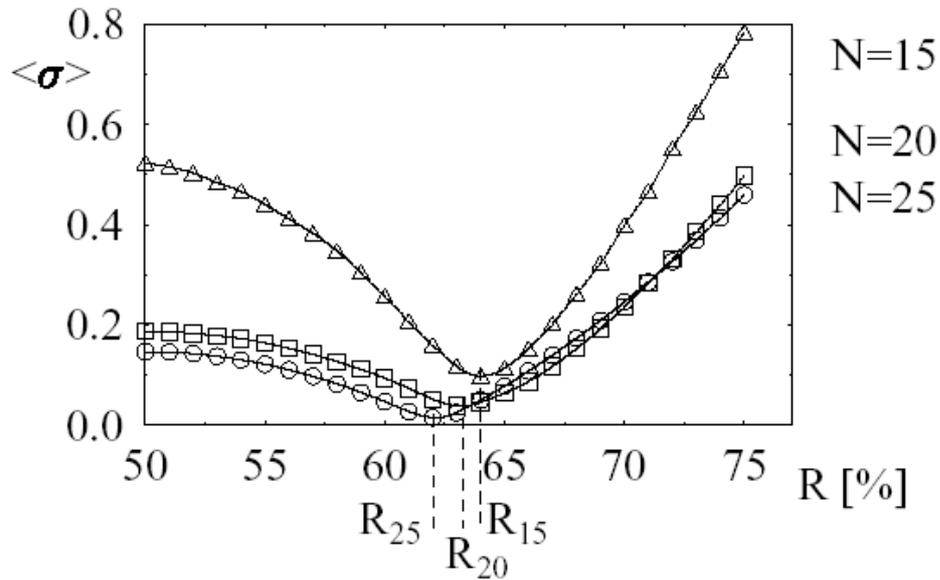

**Fig. 6. Dependence of the mean (with respect to the statistical measure) of the square root of the sum of square residuals between Banzhaf index values and voting weights (<σ>) on the value of the threshold (*R*) for *N* = 15, 20, 25.**

Exemplary computations performed for even larger 'Unions' ($N = 100$, 150, and 200), support a natural conjecture that in the thermodynamical limit, $N \rightarrow \infty$, the optimal threshold converges to 50% (see also LiM04). Note that the efficiency of such a voting system, devised for such an ensemble of randomly chosen $N$ countries, is not a free parameter anymore, but is determined by a concrete choice of $R_{opt}$. Somewhat fortunately, the value of the Coleman index $A = 16.6\%$ for the P-62 system remains at the reasonable level. However, the monotonic decrease of $R_*$ with $N$ is consistent with the well known fact that by every extension of the Union, the quota should be decreased to preserve the effectiveness of the system.

## 11. Concluding remarks

a. Viewed from the perspective of Spain's and Poland's voting power on the Council of Ministers of the European Union, the arrangements laid down in the Treaty of Nice are much more favourable than the draft of the European Constitution.

b. Neither the voting rules approved at Nice nor those adopted by the European Convention are based on objective criterions. Both systems are not representative. In particular, the system of Nice is not favourable e.g. for Germany and Latvia, while the system of the Convention handicap Spain, Poland and all middle-size states.

c. As noted by Penrose in 1954 the system of two criteria: one per capita, one per state, cannot be representative. This is just the case for the system proposed by the European Convention. It privileges the large and the small countries, at the expense of the countries of intermediate population. This general property cannot be entirely reversed by manipulating with both majority quota.

d. A single criterion system based on the square root low of Penrose is simple, representative, easily extendible, transparent, and objective: it does not favour nor handicap any European country.

e. The dependence of the voting power on the qualified majority quota displays a critical behaviour. At the quota equal to 62% of all votes there exist a critical point, at which the Penrose law is fulfilled with a very high accuracy. This quota offers a voting system with an appropriate efficiency, four times larger then this characterising the system of Nice.

f. Existence of such a critical point has been confirmed for an ensemble of several random 'Europe's', the population of all 25 'countries' has been drawn with respect to the statistical (uniform) measures on the simplex. The position of the critical point depends weakly on the concrete realisation, decreases slowly with the number $N$ of countries in the Union, and reaches 61% for $N = 28$.

g. The computation of voting power is a challenging task from both a mathematical and a computational technology point of view. Therefore, in attempting to work out an optimal compromise, it is therefore advisable to take advantage of the knowledge of experts working on the mathematical theory of voting. The examples cited in this paper show that in the matter of numbers, weights and algorithms, European politicians often take decisions that prove difficult to explain rationally later on, making their effective defence not easy.

## Acknowledgements

We wish to express out gratitude towards the co-author of our compromise solution, Marek Pieńkowski OP, for his collaboration in working it out, as well as to Erik Aurell, Frederic Bobay, Maria Ekes, Janusz Hołyst, Mikołaj Jasiński, Werner Kirsch, Andrzej Kotański, Joachim Krug, Dennis Leech, Moshé Machover, Jerzy Ombach, Tadeusz Sozański, Rafał Trzaskowski, Łukasz Turski, Aleksander Urbański, Tomasz Zastawniak, Włodzimierz Zwonek, and Michał Życzkowski for their valuable comments and suggestions.

## Appendix

### Voting systems and a mechanical toy model

To illustrate the features of the double majority voting system we are going to discuss a simple physical example from the theory of elasticity. Consider a heavy, highly elastic beam placed at the ground. Any attempt to lift it using two forces concentrated at both ends leads inevitably to a deformation: due to the internal weight of the beam its centre remains always below both ends. To reduce the deformation one may try to add some additional concentrated forces along the span. However, such an action will not allow us to reduce the deformations of the beam to zero. The only way to lift the beam without any deformation is to apply the force continuously along entire beam.

This mechanical example is designed to visualise the changes of the voting power in European Union. The elastic beam represents the set of European countries ordered according to the population. The double majority system proposed by the Convent corresponds to the two forces load. The *'per capita'* criterion, corresponds to the first force concentrated at the left end of the beam, increases the voting power of the countries with the largest population. The *'per state'* criterion corresponds to the second force, applied at the right end of the beam, and increases the voting power of the countries of the smallest population – compare Fig. 7. This is the case, independently of the relative size of the forces applied at both ends, or of the values of both quotas. They may indeed influence the overall shape of the beam, but one can not get rid of the deformation just by manipulating with their weights.

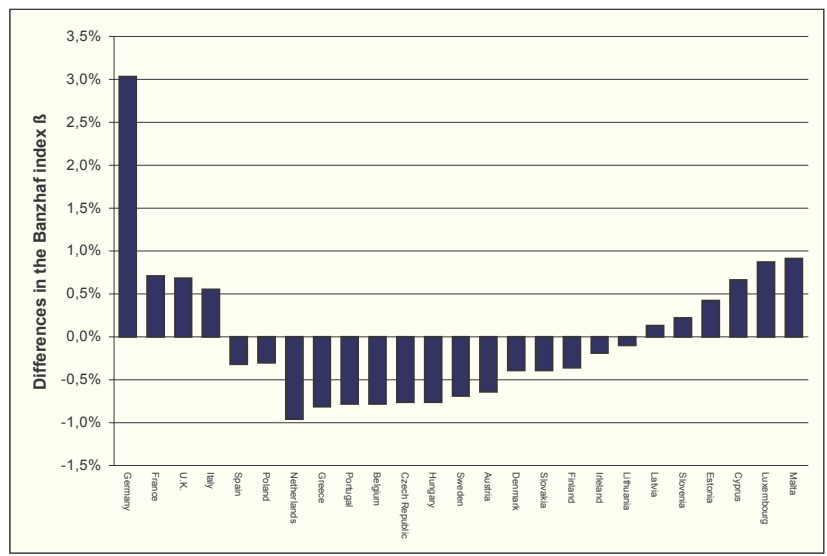

**Fig. 7. Differences in voting power in the EU Council between the proposal put forward in the Draft Constitution and proposal P-62.**

Also improving the voting system by adding ad hoc some further criterions will not make it representative (or applying the additional forces to the beam will not make it straight). This concerns also recently proposed variations of the system proposed in the Draft of the Constitution, in which the quota for populations and states are varied. Non representativeness of such systems[16] is demonstrated in Fig. 8, which shows the relative voting powers (3) of citizens of some exemplary states. The countries selected are ordered according to the population - observe that the data obtained for each of the system studied resemble a deflected beam.

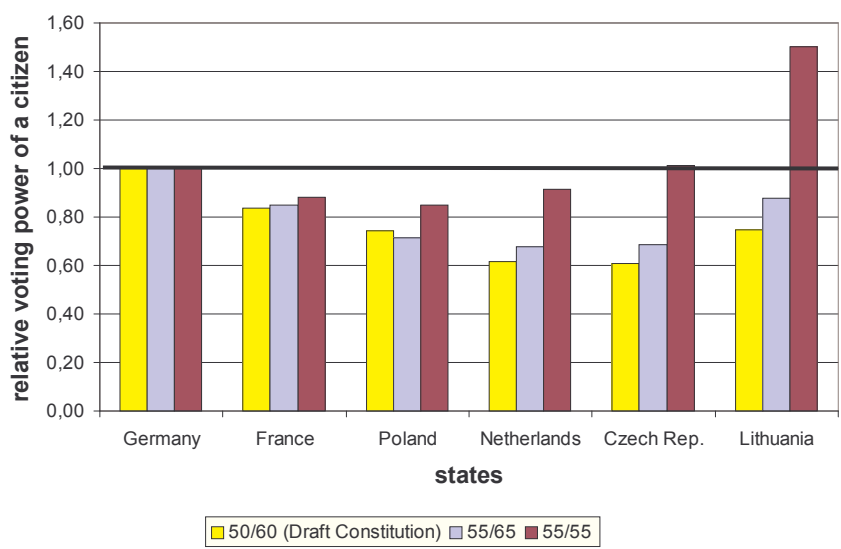

**Fig. 8. Voting power $\mu_i$ of a citizen of France, Poland, Netherlands, Czech Republic and Lithuania relative to the voting power of a citizen of Germany, computed for the Draft of Constitution 50/50 and its variants: 55/65 and 55/55.**

According to the law of Penrose, the system can be made representative, by using the single criterion based the weight proportional to the square root of the populations. In our mechanical example, this corresponds to the case of applying a uniform continuous force along the entire beam.

-----------------------------------------------------------

<sup>1</sup> His last name is also now more widely known thanks to the activities of his three exceptional sons: the physicist Oliver, mathematician Roger and the chess player Jonathan.

<sup>2</sup> On the rule of parity among the largest Member States of the Community, at a meeting held in Bonn on 4<sup>th</sup> April 1951 during the preparations leading up to the signing of the "Treaty establishing the European Coal and Steel Community". Jean Monnet, *Mémoires*, Fayard, 1976, p. 414-415 (see Bo01).

<sup>3</sup> The literature also examines the Coleman index (1971), which is directly proportional to the Banzhaf index, as well as other values, such as, for example Deegan-Packel (1979), Hoede-Bakker (1982) and Holler (1982) indices, the analysis of which allows one to draw conclusions similar to those discussed in this article. A full discussion of voting power theory may be found in the monograph by Felsenthal and Machover (see FM98). A historical perspective on the issues involved is presented in FM04a, FM04c.

<sup>4</sup> Although the Nice arrangements apply equally to Bulgaria and Romania, we only took into consideration the 25 countries which will be members of the Union following enlargement in May 2004.

<sup>5</sup> The indices have been calculated based on the population data for 1 January 2003, published by EUROSTAT [Eurostat, *EU - Total population on 1 Jan 2003*] taking into consideration all 33,554,432 possible coalitions among 25 Member States; analogous results may be found in other papers.

<sup>6</sup> It has recently been proposed to change the double-majority system proposed by the Convention and raise the population threshold from three-fifths to two-thirds. This so-called "*El Pais*" proposal is even more disadvantageous for Poland, as additionally it increases the difference between our country and the other "large" Member States of the Union. Following such a change, the value of the Banzhaf index for Poland is 6.7%, while for Germany it is 14.5%! Nearly as disadvantageous for Poland is the recently made proposal to fix both the population threshold and the number of countries at 55%. Following the adoption of such a proposal, the values of the Banzhaf index for Poland and Germany would be 5.4% and 9.4% respectively. See also BW 04.

<sup>7</sup> See Pe52, p.73.

<sup>8</sup> For the sake of simplicity, let us assume that all the citizens of a given country participate, which has no bearing on the paradox concerned.

<sup>9</sup> Let us note though that the low efficiency of the system does not at all have to be treated as a disadvantage, as it automatically forces EU Member States to seek a compromise.

<sup>10</sup> See also Pl04. Other compromise proposals were recently put forward by Paterson and Silársky (see PS03, Project *Synthesis*) and Baldwin and Widgren (see BW03b, Project *Emergency Repair*). After the publication of our proposal, an interesting modification had already been presented by an international group of students from the European College in Natolin (see COE04).

<sup>11</sup> Polish journalists, referring to our proposal, have coined for it an appealing name *Jagiellonian Compromise*.

<sup>12</sup> See end note 5.

<sup>13</sup> Previously, such an equalisation of the values of voting weights and voting power has been observed and theoretically explained in the case of a very large number of countries, which constitutes the substance of Penrose's Limit Theorem (see Pe52, LiM04). See also Sect. 10.

<sup>14</sup> The only adjustment required would involve a minor reduction in the value of the threshold set. Thus, once these countries accede to the EU, the system being described here would be described more aptly as P-61.

<sup>15</sup> Some politicians believe that any voting system on the EU Council must observe the principle that no decision may be approved by vote if it is not supported by a majority of the EU Member States (CONFER 4769/00). While the P-62 does not, in fact, satisfy this requirement, the smallest possible (and highly unlikely) winning coalition numbers the nine (biggest) countries, which together represent 82.7% of the total population of the Union. What is more, among all winning coalitions, 97.5% comprise 13 or more countries. If necessary, the P-62 system could be modified by adopting weights that are proportional to the number of citizens raised to the power of 0.53, setting the threshold at 66% of the sum of weights, and by adding the second criterion of a mandatory majority of the Member States. A system modified in this manner would be characterised by the same transparency (for all the Member States, the weight of their vote would be proportional to their respective voting power), but would be more complex than P-62.

<sup>16</sup> Similar conclusions were obtained very recently by Baldwin and Widgrén (see BW04).